\documentclass[journal,transmag]{IEEEtran}
\usepackage{color}
\usepackage[pdftex]{graphicx}

\begin{document}

\title{Spin-Singlet and Spin-Triplet Josephson Junctions\\for Cryogenic Memory}

\author{\IEEEauthorblockN{Norman O. Birge\IEEEauthorrefmark{1}
and Manuel Houzet\IEEEauthorrefmark{2}}
\IEEEauthorblockA{\IEEEauthorrefmark{1}
Michigan State University, East Lansing, MI 48824, USA}
\IEEEauthorblockA{\IEEEauthorrefmark{2}
Univ.~Grenoble  Alpes,  CEA,  IRIG-Pheliqs,  F-38000  Grenoble,  France}
\thanks{Manuscript received October 2, 2019.
Corresponding author: N. O. Birge (birge@msu.edu).}}


\IEEEtitleabstractindextext{%
\begin{abstract}
Due to the ever increasing power and cooling requirements of large-scale computing and data facilities, there is a worldwide search for low-power alternatives to CMOS.  One approach under consideration is superconducting computing based on single-flux-quantum logic.  Unfortunately, there is not yet a low-power, high-density superconducting memory technology that is fully compatible with superconducting logic.  We are working toward developing cryogenic memory based on Josephson junctions that contain two or more ferromagnetic (F) layers.  Such junctions have been demonstrated to be programmable by changing the relative direction of the F layer magnetizations. There are at least two different types of such junctions -- those that carry the innate spin-singlet supercurrent associated with the conventional superconducting electrodes, and those that convert spin-singlet to spin-triplet supercurrent in the middle of the device.  In this paper we compare the performance and requirements of the two kinds of junctions.  Whereas the spin-singlet junctions need only two ferromagnetic layers to function, the spin-triplet junctions require at least three.  In the devices demonstrated to date, the spin-singlet junctions have considerably larger critical current densities than the spin-triplet devices.  On the other hand, the spin-triplet devices have less stringent constraints on the thicknesses of the F layers, which might be beneficial in large-scale manufacturing.
\end{abstract}

}

\maketitle

\IEEEdisplaynontitleabstractindextext
\IEEEpeerreviewmaketitle

\section{Introduction}

There has been a resurgence of interest in superconducting electronics since the realization that standard CMOS-based electronics is reaching limits dictated by power requirements and heat dissipation.  Large-scale computing installations and data centers already use several percent of the world's electricity production \cite{Marashi2019}, and that share is predicted to grow significantly over the next decades.  Due to the recent development of highly energy-efficient superconducting logic based on the manipulation of single magnetic flux quanta (SFQ) \cite{Mukhanov2011}, superconducting computing appears to be a promising solution to the energy problem \cite{Holmes2013, Manheimer2015}.

Building a superconducting computer requires more than logic, however.  One of the largest barriers to the development of large-scale superconducting computer is memory \cite{NSA-Tech-Assessment2005}.  The native memory associated with SFQ logic works well, but does not have high enough density to serve as the main memory in a large-scale computer.  Several alternatives have been proposed \cite{Soloviev2017}.  This paper focuses on a memory concept developed by workers at Northrop Grumman Corporation called Josephson Magnetic Random Access Memory or JMRAM \cite{Dayton2018}.  The active element in a JMRAM memory cell is a Josephson junction containing two or more ferromagnetic layers, whose characteristics depend on the magnetic configuration of those layers.  From the magnetic point of view, JMRAM is reminiscent of traditional magnetic random access memory (MRAM); the readout, however is accomplished using a superconducting circuit.  While the original version of JMRAM relied on controlling the magnitude of the critical current in the active element, more recent versions rely on controlling the ground-state phase shift across the junction to be either $0$ or $\pi$.  (We do not consider so-called ``$\phi_0$-junctions" here, which can have an arbitrary ground-state phase difference across the terminals.)  This digital aspect of the readout provides greater margins for the properties of the active element, hence more robust performance of a large-scale memory array \cite{Dayton2018}.

There are two different types of magnetic Josephson junction that can have a controllable phase state.  The first type carries the innate spin-singlet supercurrent expected in any Josephson junction that uses conventional superconductors for the electrodes.  Golubov \textit{et al.} showed in 2002 that incorporating two ferromagnetic (F) layers into such a junction would allow the junction to be either in the $0$ state or the $\pi$ state depending on the relative orientation of the two F-layer magnetizations \cite{Golubov2002}.  The second type of controllable junction converts spin-singlet electron pairs to spin-triplet pairs inside the junction.  The possibility of generating such spin-triplet pairs in S/F hybrid systems was first pointed out in \cite{Bergeret2001}, where it was shown that the key ingredient is the presence of non-collinear magnetizations in the system.  Later, Ref.~\cite{Houzet2007} suggested a practical Josephson junction architecture for realizing the prediction of \cite{Bergeret2001}. In this proposal, the junction contains three magnetic layers, with the magnetizations of adjacent layers non-collinear. The ground-state phase difference across the junction depends on the relative orientations of all three F-layer magnetizations. In the last several years, demonstrations of magnetically-controlled ground state phases have been carried out for both types of junctions \cite{Gingrich2016, Glick2018, Madden2019}.

The purpose of this paper is to compare the performance and requirements of spin-singlet and spin-triplet phase-controllable Josephson junctions, cf. setups in Figs. 1(a) and 2(a).  The next two sections briefly describe the physics of the two kinds of devices.  Section IV then presents ``phase diagrams'' for both types of junctions, showing how the ground-state phases of the different magnetic states depend on the thicknesses of the two F layers in spin-singlet devices or of the outer two F layers in spin-triplet devices.  The paper ends with a brief discussion of future work needed to bring these devices to the level of practical application.

\section{Spin-Valve Josephson Junctions}

The physics of the proximity effect in S/F hybrid systems has been explained in several review articles~\cite{Buzdin2005, Bergeret2005, Eschrig2011, Linder2015}.  The basic idea can be summarized as follows: when a spin-singlet Cooper pair diffuses across the interface from a conventional S material to an adjacent F material, the two electrons that make up the pair are no longer bound to each other.  In S, the two electrons have opposite momenta, i.e. they come from opposite sides of the Fermi surface.  In F, however, the majority (spin-up) and minority (spin-down) electron bands are offset from each other in energy, so the two electrons with energies at the Fermi surface cannot find states with exactly opposite momenta.  Instead, they acquire a finite momentum $Q = k_{F\uparrow} - k_{F\downarrow} \approx 2 E_{ex}/v_F$, where $k_{F\uparrow}$ and $k_{F\downarrow}$ are the Fermi momenta for each spin, $E_{ex}$ is the exchange energy in F in the parabolic band approximation, and $v_F$ is the Fermi velocity. Alternatively one can say that the pair correlation function displays oscillations in F with wavevector $Q$.  In a one-dimensional system with no extrinsic sources of dephasing, the oscillations in the pair correlation function would persist over a distance comparable to the ``normal-metal coherence length'', $\xi_N = \hbar v_F/k_BT$, where $T$ is the temperature.  In three-dimensions, interference between the oscillations in different spatial directions leads to an algebraic decay of the oscillations. In a diffusive system, the oscillations decay exponentially on the scale $\xi_F = \sqrt{\hbar D/E_{ex}}$, where $D$ is the diffusion constant. Strong F materials have large values of $E_{ex}$, of order of 1 eV, hence $\xi_F$ in such materials is typically less than 1 nm \cite{Robinson2006}. Weak F alloys can have $\xi_F$ as long as several nm, which eased the requirements for the first experimental groups who observed signatures of the pair correlation oscillations in S/F systems \cite{Kontos2001,Ryazanov2001}.

The oscillations in the pair correlation function discussed above lead to oscillations in the ground-state phase difference across an S/F/S Josephson junction \cite{Buzdin1982}.  A simple ``hand-waving'' derivation of that fact was presented by Eschrig in 2011 \cite{Eschrig2011}.  For completeness, we reproduce his argument here.  Since S is assumed to be a conventional s-wave, spin-singlet superconductor, the Cooper pair spin state is the singlet:
\begin{equation}
\label{eq:S}
|\chi_S\rangle=|s=0,m_s=0\rangle \equiv \frac 1{\sqrt{2}} \left(|\uparrow\downarrow\rangle-|\downarrow\uparrow\rangle\right),
\end{equation}
where $s$ is the total spin of the pair and $m_s$ is the spin projection along the magnetization axis. As a pair propagates a distance $X$ through F, each component in (1) acquires an oscillatory factor equal to $\exp(\pm iQX)$, with the sign depending on whether the up-spin electron or the down-spin electron has the larger Fermi momentum.  The wavefunction of the pair thus becomes
\begin{equation}
\label{eq:F}
|\chi_F(X)\rangle=\frac 1{\sqrt{2}}\left(e^{iQX}|\uparrow\downarrow\rangle-e^{-iQX}|\downarrow\uparrow\rangle\right).
\end{equation}
Finally, we re-write (\ref{eq:F}) using the Euler formula to obtain
\begin{equation}
\label{eq:Fbis}
|\chi_F(X)\rangle=\cos(QX)|0,0\rangle+i\sin(QX)|1,0\rangle
\end{equation}
with $|1,0\rangle\equiv \left(1/{\sqrt{2}}\right)\left(|\uparrow\downarrow\rangle+|\downarrow\uparrow\rangle\right)$.  The far S electrode accepts only spin-singlet electron pairs, hence the Josephson coupling energy of the junction, and the supercurrent through it, are proportional to
\begin{equation}
\label{eq:EJ}
E_J(d_F)\propto \langle 0,0|\chi_F(d_F)\rangle=\cos(d_F/\xi_F),
\end{equation}
where we have used the fact that $\xi_F=Q^{-1}$ in the ballistic limit.  In this simple model, $0-\pi$ transitions occur at $d_F/\xi_F \approx (n + 1/2)\pi$, for $n = 0, 1, 2,\dots$. A negative Josephson energy in (\ref{eq:EJ}) corresponds to the junction being in the $\pi$ phase state.

The preceding ``derivation'' does not provide quantitative information about the magnitude of the supercurrent.  Not only is it restricted to purely 1D ballistic systems, but it neglects reflections at both S/F interfaces due to Fermi surface mismatch.  Nevertheless, it provides a correct understanding of the oscillations between $0$-junctions and $\pi$-junctions as the F-layer thickness varies.  In a 3D system, one must sum pair trajectories that leave the interface at all angles \cite{Buzdin1982}.  The result of the calculation (valid only near the superconducting critical temperature $T_c$) is
\begin{equation}
\label{eq:EJ-3D}
E_J(d_F)\propto \left(\frac{d_F}{\xi_F}\right)^2\int_{d_F/\xi_F}^\infty\frac{\cos y}{y^3}dy.
\end{equation}
For large $d_F/\xi_F$, Eq.~(\ref{eq:EJ-3D}) is nearly equal to $-\sin(d_F/\xi_F)/(d_F/\xi_F)$, with consequent $0-\pi$ transitions at $d_F/\xi_F \approx n\pi$, but that asymptotic form misses the first $0-\pi$ transition given by Eq.~(\ref{eq:EJ-3D}), which occurs at $d_F/\xi_F\approx 0.33\pi$.  In the diffusive limit, one must solve the Usadel equations to calculate $E_J$~\cite{Buzdin1991}. The result of that calculation in the simplest limit neglecting spin-flip and spin-orbit scattering is that $0-\pi$ transitions occur at thicknesses $d_F/\xi_F \approx (n + 3/4)\pi$.

An S/F/S Josephson junction containing only a single F layer has a set phase difference; i.e. it is not controllable.  (The exception is when the ferromagnetic in F is very weak, where the temperature can be used to control the phase state of a junction very close to the $0-\pi$ transition \cite{Ryazanov2001}.)  In 2002, Golubov \textit{et al.} (see above) showed that adding a second F layer enables one to control the phase state of the junction \cite{Golubov2002}.  Shortly thereafter, that possibility was suggested in an experimental paper by Bell \textit{et al.}~\cite{Bell2004}, but those authors (and others later \cite{Baek2014,Qadar2014}) demonstrated control only of the critical current amplitude, not the phase.  The operating principle of the phase-controllable junction is that, if the magnetizations of the two F layers are parallel (P state), then their influence on the oscillations of electron pairs is cumulative, i.e. the accumulated phase shift of the pair is
\begin{equation}
\label{eq:phiP}
\phi_P=\frac{d_{F1}}{\xi_{F1}}+\frac{d_{F2}}{\xi_{F2}},
\end{equation}
where $d_{F1}$ and $d_{F2}$ are the thicknesses of the two F layers, and $\xi_{F1}$ and $\xi_{F2}$ are the two ``ferromagnetic correlation lengths'' for pairs.  When the magnetizations are antiparallel (AP state), then the majority and minority bands switch roles in the second F layer, so that the total electron pair phase shift is:
\begin{equation}
\label{eq:phiAP}
\phi_{AP}=\frac{d_{F1}}{\xi_{F1}}-\frac{d_{F2}}{\xi_{F2}},
\end{equation}

In the simplest 1D ballistic model, making a phase-controllable junction requires that one choose the thicknesses $d_{F1}$ and $d_{F2}$ in such a way that $\cos\phi_P$ and $\cos\phi_{AP}$ have opposite sign.  The values of $\phi_P$ and $\phi_{AP}$ that give rise to 0 or $\pi$ junctions in 3D diffusive systems are modified somewhat, as discussed above.

\section{Spin-Triplet Josephson junctions}

To understand how spin-singlet pairs can be converted to spin-triplet pairs inside a Josephson junction, we start from Eq.~(\ref{eq:Fbis}), which showed how the $m_s=0$ triplet component of the pair amplitude was generated by a single F layer adjacent to one of the S electrodes. Let us set the direction of magnetization of the first F layer (now called F1) to be along the $z$ axis. If we add a second F layer, called F2,  next to the first, with its magnetization rotated by an angle $\theta_{12}$ around the $y$-axis, then the $m_s=0$ triplet component generated by F1 transforms into all three triplet components in the new basis.  So the wavefunction in F2 takes the form:
\begin{eqnarray}
\nonumber
|\chi_{F2}\rangle&=&\cos\left(\frac{d_{F1}}{\xi_{F1}}\right)|0,0\rangle
\\
&&+i\sin\left(\frac{d_{F1}}{\xi_{F1}}\right)\left(\frac {\sin\theta_{12}}{\sqrt{2}}|1,-1\rangle_{\theta_{12}}\right.
\nonumber
\\
&&\left.+\cos\theta_{12} |1,0\rangle_{\theta_{12}}
-\frac {\sin\theta_{12}}{\sqrt{2}}|1,1\rangle_{\theta_{12}}\right),
\label{eq:Fter}
\end{eqnarray}
where we have used the subscripted angle $\theta_{12}$ to indicate the new, rotated, spin basis. The special feature of spin-triplet junctions is that the $m_s=\pm1$ triplet components (often called ``equal-spin triplets'' in the literature) are long-ranged in F2, because the two electrons are in the same spin band.  In addition, those two components do not oscillate in space the way the singlet and $m_s=0$ triplet components do.  If we make F2 much thicker than its ferromagnetic coherence length, i.e. $d_{F2}\gg\xi_{F2}$, then the spin-singlet and $m_s=0$ triplet components decay to negligible values, so we are left only with the two equal-spin triplet components.

By inserting a third F layer (called F3) in-between F2 and another superconductor, it is possible to convert the long range triplet pairs back into singlet Cooper pairs, provided that the magnetization in F3 is rotated by an angle $\theta_{23}$ with respect to the magnetization of F2. By the same argument as the one used to derive Eq.~(\ref{eq:Fter}), we infer that the dependence of the Josephson coupling energy on the thicknesses of F1 and F3 and the rotation angles $\theta_{12}$ and $\theta_{23}$ is given by
\begin{equation}
\label{eq:EJ-triplet}
E_J\propto -\sin(d_{F1}/\xi_{F1})\sin(d_{F3}/\xi_{F3})\sin \theta_{12}\sin\theta_{23}.
\end{equation}
The advantage of the spin-triplet over spin-valve Josephson junctions is that the $0-\pi$ transition is achieved simply by reversing the magnetization of one of the outside layers, F1 or F3, regardless of the thicknesses of those layers.

\section{Phase diagrams of Spin-Singlet and Spin-Triplet Phase Controllable Josephson Junctions}

We can now present a rudimentary phase diagram for both spin-singlet and spin-triplet junctions.  We present results valid in the diffusive limit, which are obtained by solving the Usadel equations.  We neglect spin-orbit and spin-flip scattering, which complicate the calculations considerably \cite{Faure2006}. We assume that the magnetic layers have the same exchange field and normal-state conductivity. We also neglect the effects of the nonmagnetic spacer layers used to decouple the magnetizations of adjacent F layers, and which are known to shift slightly the positions of the 0-$\pi$ transitions \cite{Heim2015}. Finally, our calculations are carried near $T_c$ and assume the so-called ``short-junction limit'', corresponding to the junction length shorter than the diffusive normal coherence length $\xi_N=\sqrt{\hbar D/k_BT}$. The main goal here is to show how the behavior of the Josephson junctions depend on the thicknesses of the ferromagnetic layers inside the junction.

For spin-singlet junctions, we use the result derived by Crouzy \textit{et al.} \cite{Crouzy2007}, which was the first calculation of the supercurrent of an S/F1/F2/S spin-valve Josephson junction in the parallel and antiparallel magnetic states for arbitrary thicknesses of F1 and F2.  (An earlier calculation by Blanter and Hekking considered only equal thicknesses of F1 and F2 in the diffusive limit \cite{Blanter2004}.)  Namely, the critical currents in the P and AP configurations are
\begin{equation}
I_c^P=I_{c0}\mathrm{Re}\left[\frac{(1+i) d_{+}/\xi_F}{\sinh(1+i) d_+/\xi_F}\right]
\end{equation}
and
\begin{equation}
I_c^{AP}=I_{c0}\mathrm{Re}\left[ \frac{2 d_+/\xi_F}{\sin(d_++id_-)/\xi_F) +\sinh(d_+-id_-)/\xi_F)}\right],
\end{equation}
respectively. Here $d_\pm=d_{F1}\pm d_{F2}$, $I_{c0}=\pi G \Delta^2/(4ek_B T_c)$ is the critical current of the junction in the absence of exchange field, with normal state conductance $G$, and $\Delta\approx 3.06 k_B\sqrt{T_c(T_c-T)}$ is the superconducting gap  near $T_c$. Figure 1(b) shows the results, and is similar to Fig. 2 in \cite{Crouzy2007} and to Fig. 1(a) in \cite{Birge2018}.  As $d_{F1}$ and $d_{F2}$ vary, $I_c^P$ and $I_c^{AP}$ vanish and reverse their signs across the thick purple and black lines, respectively. Each area delimited by those thick lines encloses a region where their product, shown in Fig. 1(b), has a definite sign.  Only regions where the signs are different (regions in red), so that P and AP configurations correspond to different phase states, are suitable for memory.  So half of the available phase space is worthless for any practical application.  In \cite{Birge2018} we showed heuristically where the existing demonstrations of phase-controllable junctions appear on the phase diagram. Figure 1(c) shows the thickness dependence of $I_c^P$ and $I_c^{AP}$ for a bilayer with equal thicknesses, $d_{F1}=d_{F2}$, i.e. along the green line in Figure 1(b).

\begin{figure}[!t]
\centering
(a) \includegraphics[width=2.5in]{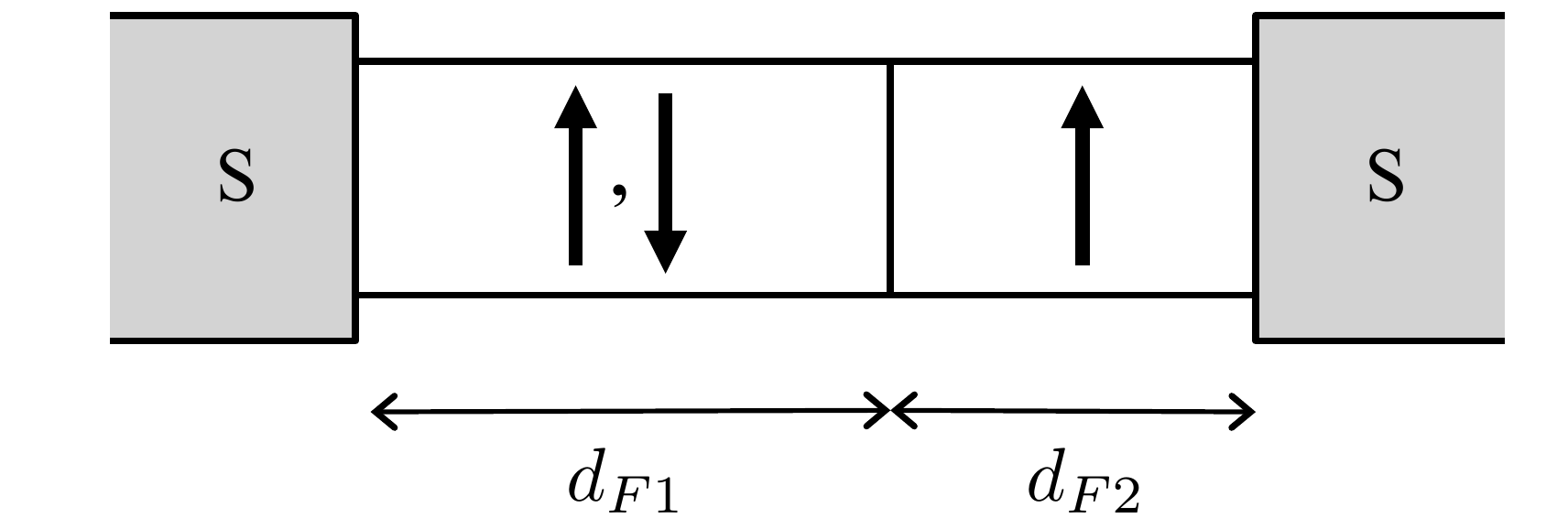}
\\
(b) \includegraphics[width=3in]{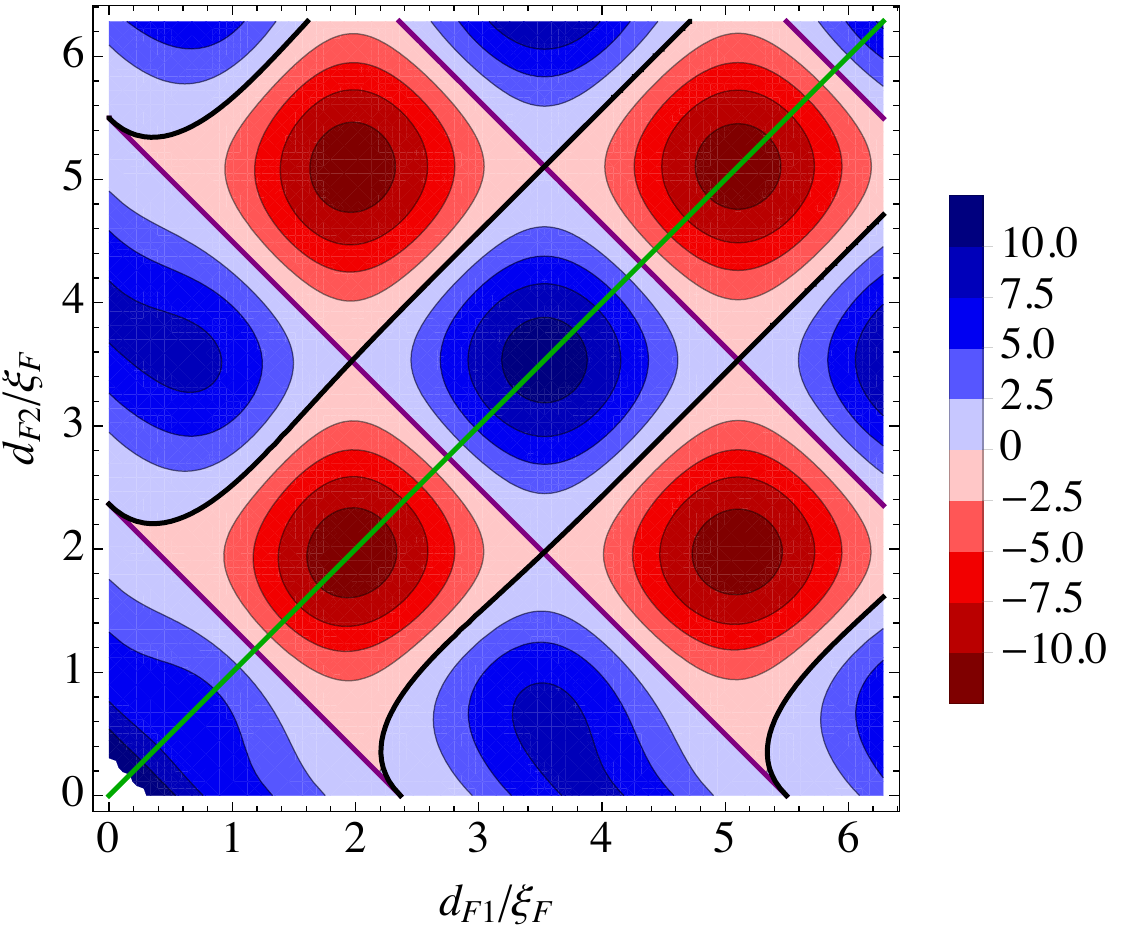}
\\
(c)
\includegraphics[width=2.5in]{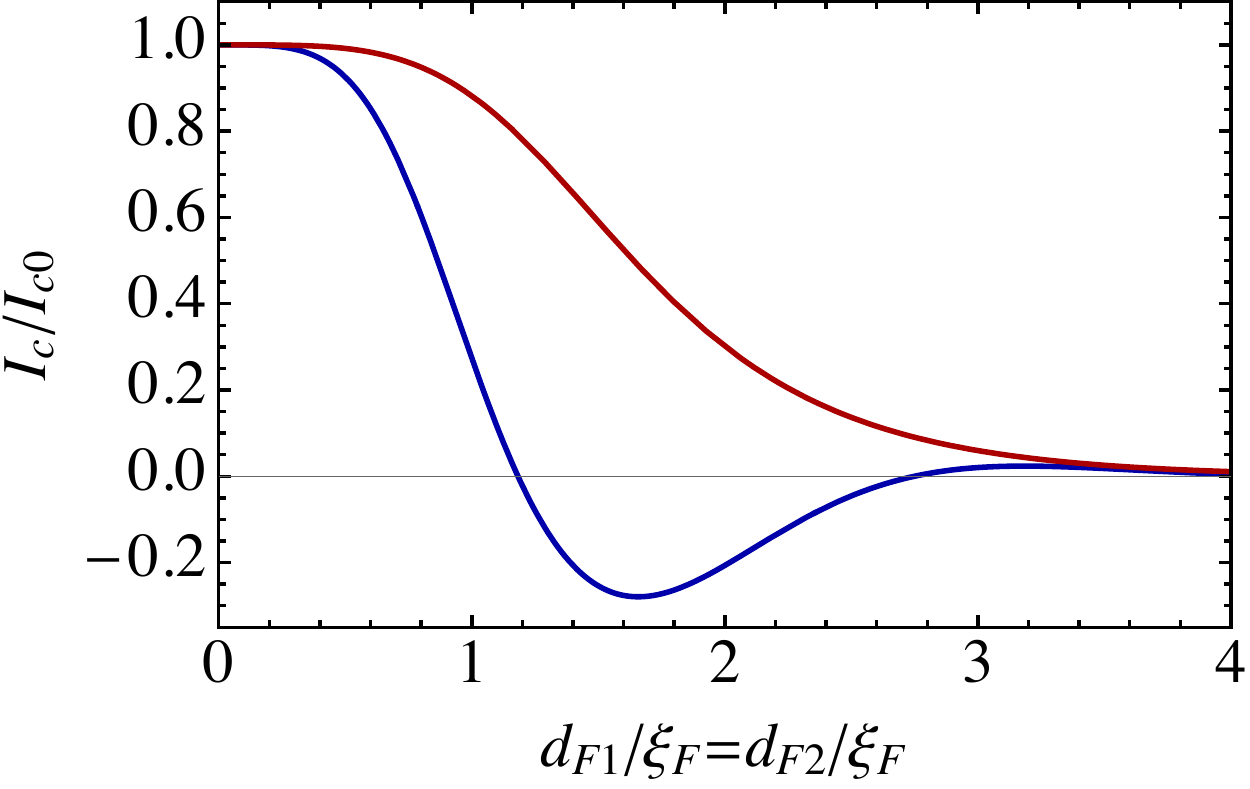}
 \caption{(a) Setup of a spin-singlet Josephson junction through a ferromagnetic bilayer. The magnetizations in each layer can be in parallel (P) or antiparallel (AP) configurations. (b) Contour plot of the product of the critical currents in P and AP configurations, $I_c^PI_c^{AP}$ (in units of $\{I_{c0}(d_{F1}+d_{F_2})/\xi_F\exp[-(d_{F1}+d_{F_2})/\xi_F]\}^2$), as a function of the thicknesses of each layer. The setup is expected to work as a cryogenic memory for thicknesses such that $I_c^P$ and $I_c^{AP}$ have opposite signs; this corresponds to the red areas delimited by the purple lines, along which $I_c^P$ vanishes, and the black lines, along which $I_c^{AP}$ vanishes. (c) Dependence of $I_c^P$ (blue line) and $I_c^{AP}$ (red line) as a function of the thickness in a bilayer with $d_{F1}=d_{F2}$ [green line in b)].}
\label{fig_sim}
\end{figure}

Our results for spin-triplet junctions extend those presented in \cite{Houzet2007} by considering the critical current through a trilayer ferromagnetic junction with unequal thicknesses $d_{F1}$ and $d_{F3}$ of the external layers. Furthermore, the central layer is assumed to be thick, $d_{F2}\gg\xi_F$, so that the spin-singlet contribution to the critical current is absent. The critical current in the P configuration of Fig. 2(a) is
\begin{equation}
I_c=4I_{c0} f\left(\frac{d_{F1}}{\xi_F}\right) f\left(\frac{d_{F3}}{\xi_F}\right)
\end{equation}
while it is reversed in the AP configuration. Here
\begin{equation}
f(x)=\frac{( x+\frac12) (\sin x \mathrm{ch}\, x-\mathrm{sh}\, x \cos x)+ (x+1)\mathrm{sh}\, x  \sin x)}{( x+\frac12) (\sin 2 x+\mathrm{sh}\, 2 x)+ (x+1) (\cos 2 x+ \mathrm{ch}\, 2 x)}
\end{equation}
with asymptotes $f(x)=x^2/2$ at $x\ll 1$ and $f(x)=(\sin x-\frac 12\cos x)e^{-x}$ at $x\gg 1$. [Note that for an arbitrary configuration of the magnetizations, the angular dependence is in general more complicated that the one given in Eq.~(\ref{eq:EJ-triplet}).] As $d_{F1}$ and $d_{F3}$ vary, the critical current changes its sign, vanishing along the thick black lines in Fig. 2(b). In contrast with the spin-singlet junctions however, the critical currents in P and AP configuration are always opposite, corresponding to different phase states, for any thickness. Figure 2(c) shows an example of the thickness dependence of the critical current in a junction with equal thicknesses $d_{F1}=d_{F3}$.

\begin{figure}[!t]
\centering
(a) \includegraphics[width=2.5in]{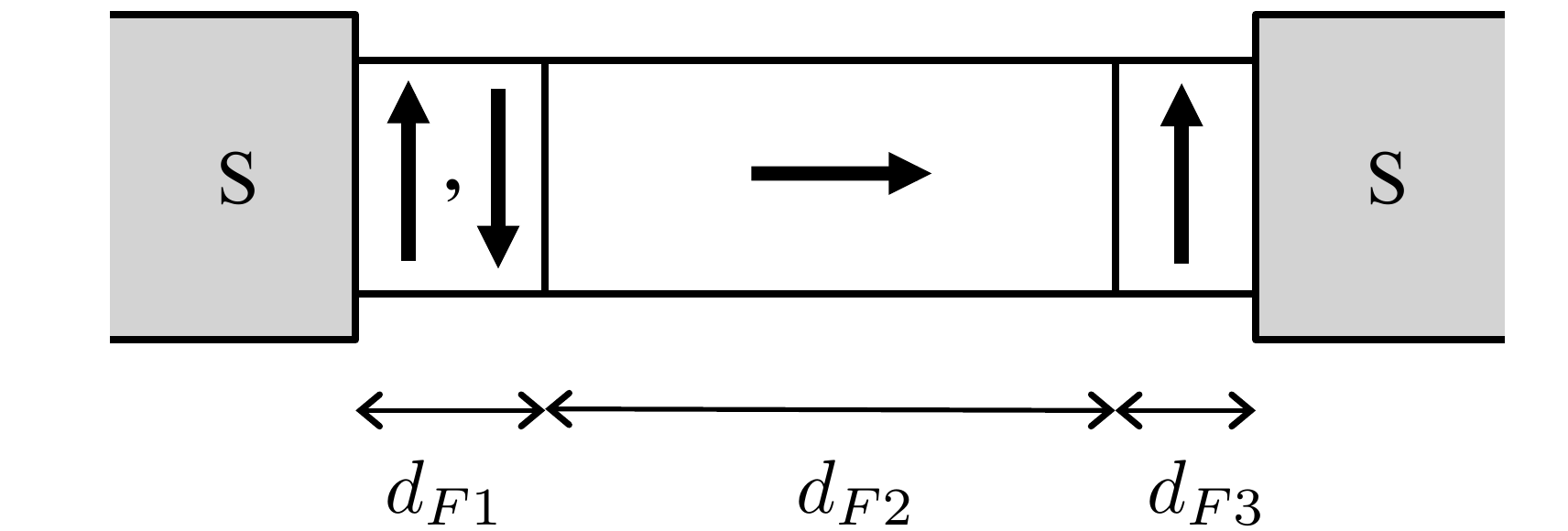}
\\
(b) \includegraphics[width=2.8in]{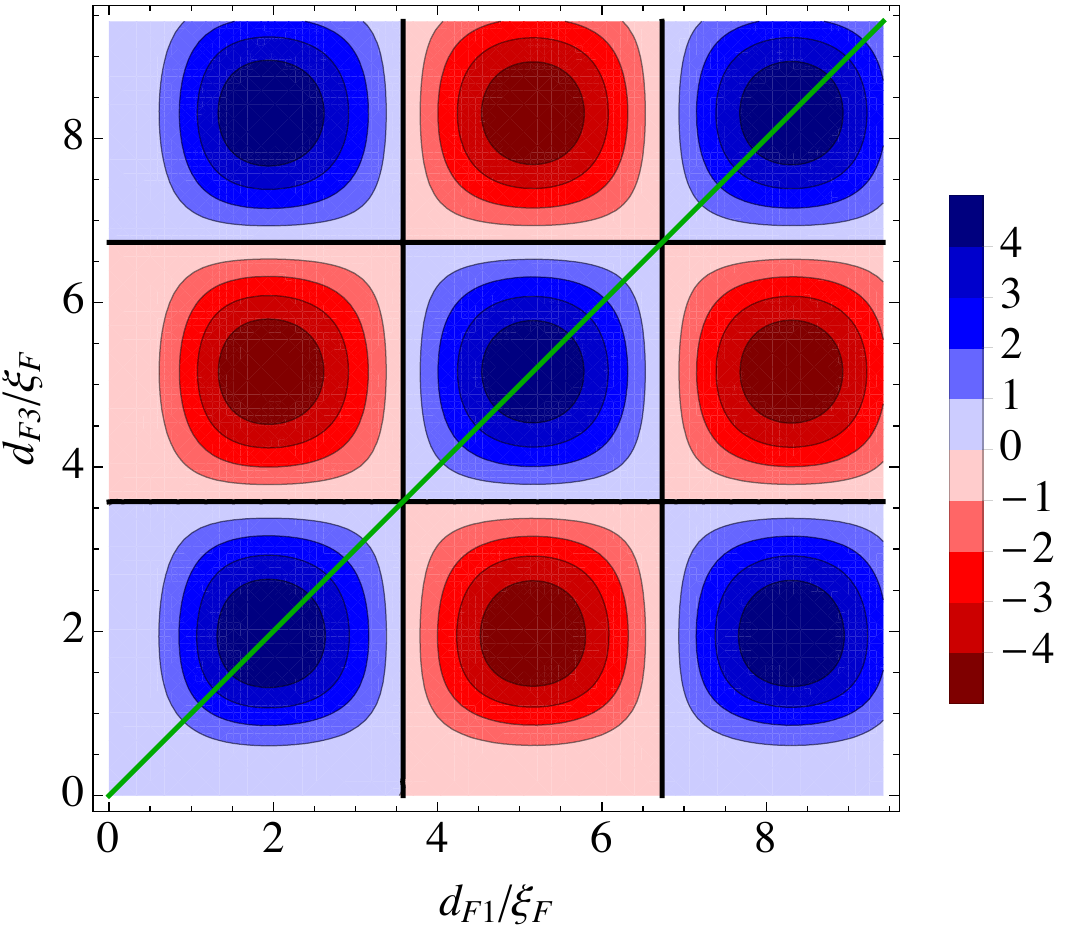}
\\
(c)
\includegraphics[width=2.6in]{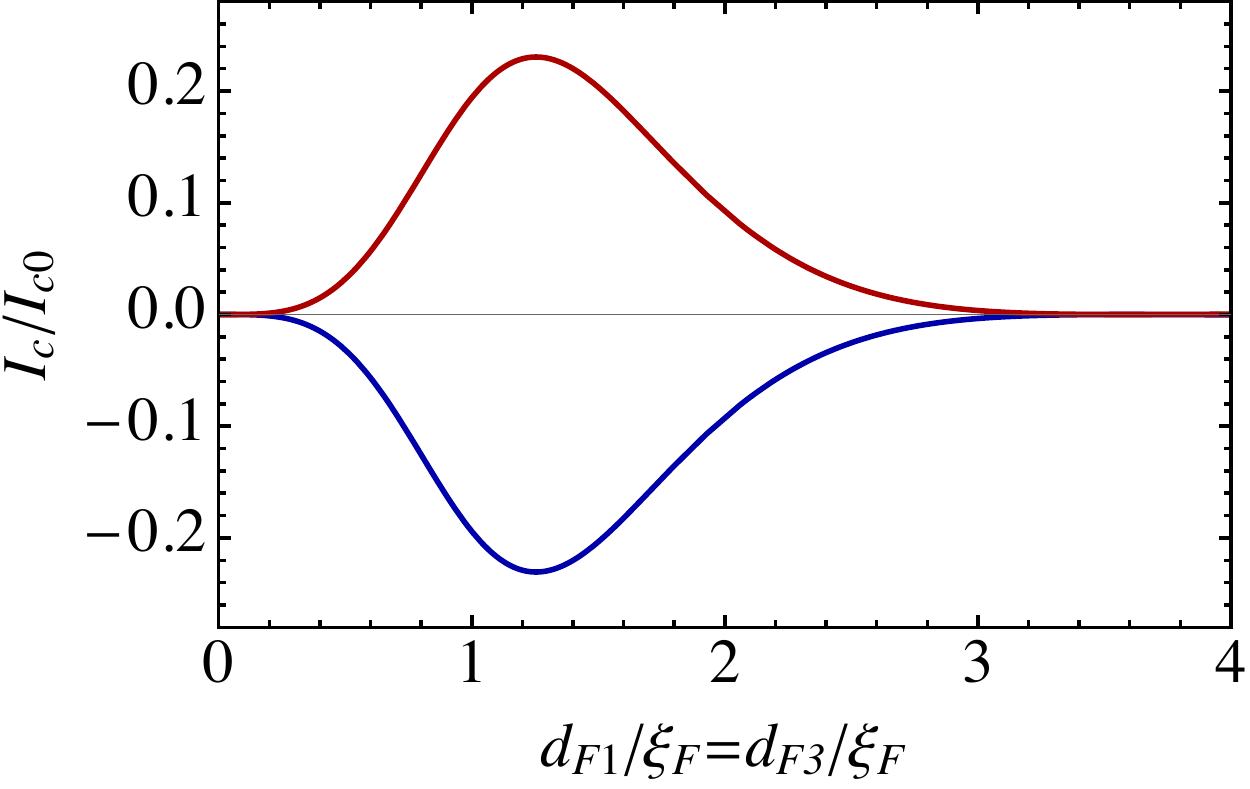}
 \caption{(a) Setup of a spin-triplet Josephson junction through a ferromagnetic trilayer with a long central layer, $d_{F2}\gg \xi_F$. The magnetizations in the external layers are perpendicular to the central one, they can be in parallel (P) or antiparallel (AP) configurations. (b) Contour plot of the critical current in P configuration, $I_c$ (in units of $I_{c0}\exp[-(d_{F1}+d_{F3})/\xi_F]$), as a function of the thicknesses of each layer. The critical current in the AP configuration has the same amplitude and opposite sign. Therefore, the setup is expected to work as a cryogenic memory for all thicknesses except for those in the vicinity of the black lines along which the critical current vanishes. (c) Dependence of $I_c^P$ (blue line) and $I_c^{AP}$ (red line) as a function of the thickness in a trilayer with $d_{F1}=d_{F3}$ [green line in b)].}
\label{fig_sim}
\end{figure}

\section{Conclusion}
For application in a large-scale memory, the spin-valve junctions have the obvious advantage that they contain only two F layers rather than three.  But the thicknesses of those layers must be chosen judiciously to avoid the blue regions in Figure 1(b).  The spin-triplet junctions work for all F-layer thicknesses (except near the lines delineating the regions in Figure 2(b)), and they produce supercurrents with equal amplitudes in the 0 and $\pi$ states.  In the demonstrations published to date \cite{Gingrich2016, Glick2018, Madden2019}, the spin-valve junctions had much larger critical currents than the spin-triplet junctions, but we are currently working on ways to increase the latter.  In both cases it will be crucial to find magnetic materials that provide consistent switching fields across many memory cells.  Only time will tell which of these approaches (if either) will lead to the successful manufacture of large-scale cryogenic memory.

\section*{Acknowledgment}

NOB thanks all the authors of \cite{Gingrich2016, Glick2018, Madden2019, Birge2018}, as well as A. Herr and N. Rizzo.  This research was partly supported by the ODNI, IARPA, via ARO contract number W911NF-14-C-0115. The views and conclusions contained herein are those of the authors and should not be interpreted as representing the official policies or endorsements, either expressed or implied, of the ODNI, IARPA, or the U.S. Government.

\end{document}